# A statistical resolution measure of fluorescence microscopy with finite photons

**Authors:** Yilun Li[1], Fang Huang[1,2,3]*

**Affiliations:**

[1]Weldon School of Biomedical Engineering, Purdue University, West Lafayette, IN, USA

[2]Purdue Institute for Integrative Neuroscience, Purdue University, West Lafayette, IN, USA

[3]Purdue Institute of Inflammation, Immunology and Infectious Disease, Purdue University, West Lafayette, IN, USA

*Corresponding author. Email: fanghuang@purdue.edu

**Abstract:** First discovered by Ernest Abbe in 1873, the resolution limit of a far-field microscope is considered determined by the numerical aperture and wavelength of light, approximately $\frac{\lambda}{2NA}$. With the advent of modern fluorescence microscopy and nanoscopy methods over the last century, it is recognized that Abbe's resolution definition alone could not solely characterize the resolving power of the microscope system. To determine the practical resolution of a fluorescence microscope, photon noise remains one essential factor yet to be incorporated in a statistics-based theoretical framework. Techniques such as confocal allow trading photon noise in gaining its resolution limit, which may increase or worsen the resolvability towards fluorescently tagged targets. Proposed as a theoretical measure of fluorescence microscopes' resolving power with finite photons, we quantify the resolvability of periodic structures in fluorescence microscopy systems considering both the diffraction limit and photon statistics. Using the Cramer-Rao Lower Bound of a parametric target, the resulting precision lower bound establishes a practical measure of the theoretical resolving power for various modern fluorescence microscopy methods in the presence of noise.

**One-Sentence Summary:** The resolving power of a far-field fluorescence microscope is determined by its diffraction limit only when infinite photons are considered. To determine the resolvability of a specimen, photon noise, however, remains one essential factor yet to be considered statistically. We proposed an information-based resolution measure quantifying the theoretical resolving power of a fluorescence microscope in the condition of finite photons. The developed approach allows us to quantify the practical resolution limit of varieties of fluorescence and super-resolution microscopy modalities, and further for microscopy developments, may be used to predict the achievable resolution of a microscopy design at various photon levels.

## Introduction

In 1873, Abbe published his work stating that microscopy resolution solely depends on the numerical aperture and wavelength of the light (*1*, *2*), a statement later verified theoretically (*3*, *4*). This limit is often sufficient for traditional microscopes where transmission, reflection, or scattered light is collected as signals (*5*), for which the signal-to-noise ratio can be increased by adjusting the illumination power. However, for fluorescence microscopes, photons—the sole source of molecular information generated by individual fluorescent probes—are limited due to the photobleaching and photochemical environment of the fluorophores (*6*, *7*).

Due to the discrete nature of light, the photon counting noise is inherent following Poisson distribution, where the signal-to-noise ratio will decrease with a decreasing number of detected photons. As a result, at low photon levels, it would be difficult to determine whether the captured intensity differences are from underlying structure or noise fluctuations, thus preventing microscopy methods from achieving their theoretical resolution limit (*8–12*).

In search of a practical resolution, researchers have attempted to create a measure based on the channel capacity in Shannon's information theorem (*13*), where noise can be included (*14–17*). The general concept is that the channel capacity of the microscope system indicates its resolving power and can be calculated from the noise level and the systems' configurations. While these works provide the theoretical relationship between the channel capacity of the microscope and signal-to-noise (SNR) ratio, numerical aperture, etc., achieving high resolution requires specific information encoding-decoding scheme. Moreover, the noise is assumed to be (frequency) band limited, which photon noise does not satisfy.

Another approach to quantify noise effect on the resolution is by evaluating the noisy images directly (*8*, *9*). For instance, by quantifying the contrast between two closely spaced point objects, Stelzer (*8*) discussed the noise influence on practical resolution in confocal and wide-field systems, which represents one of the first demonstrations of the critical difference between practical resolvability in the presence of noise and Abbe's resolution limit. However, these empirical methods rely on existing microscopy data and well-controlled specimens. To date, a theoretical resolution measure incorporating photon statistics remains missing.

Here, we propose a quantitative and theoretical measure of the resolving power of microscopes considering numerical aperture, emission wavelength, and photon statistics. Specifically, we consider the grating's phase estimation precision as a measure of the resolving power of the imaging system. Based on an adjustable criterion of the achievable precision lower bound, we defined the information-based resolution (IbR) and used it in evaluating and distinguishing the significant resolving power differences among various conventional and super-resolution imaging modalities including wide-field microscopy, confocal microscopy (*18*, *19*), structured illumination microscopy (SIM) (*20*, *21*), and image scanning microscopy (ISM) (*22–24*). We expect IbR to be a useful measure in estimating the noise-considered resolution to guide and validate the design of newly developed or proposed imaging modalities.

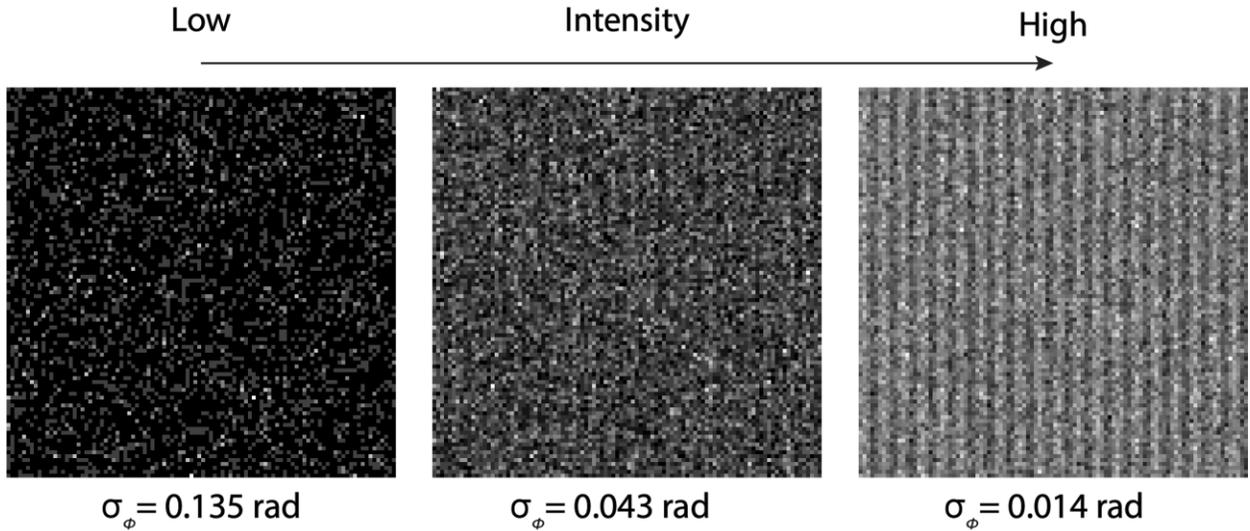

Fig. 1. Visualization of noise influence on the resolvability of a single-tone sinusoidal wave grating object. The resolvability in terms of $\sigma_\phi$ is considered the resolving power of this wide-field system. Noisy images are simulated with different intensities in this wide-field system. The frequency of the object is $\frac{1.6NA}{\lambda}$. Figures from left to right each has photon density of 500 photons/μm², 5000 photons/μm², and 50,000 photons/μm². $\sigma_\phi$ is the phase estimation precision limit of the sinusoidal object. Images are simulated under the condition: field of view (FOV) 10 μm×10 μm, $NA = 1.4$, wavelength $\lambda = 0.7\ \mu m$, immersion media refractive index $n = 1.5$, and camera pixel size 0.1 μm.

## Results

### *Model*

Inspired by Abbe's experimental approach based on the visibility of the gratings under microscopes (*1*), we examine the achievable phase estimation uncertainty of theoretical 2D sinusoidal patterns under microscopes through Fisher information (*25*, *26*).

Consider a 2D grating object expressed as

$$obj(x, y) = U_{ave}[1 + \alpha \sin(2\pi k l x + \phi)]. \tag{1}$$

Here $(x, y)$ are pixel index in two dimensions, $U_{ave}$ is the average photon count per pixel, $l$ is the pixel length, $\alpha$ is the relative amplitude, $k$ is the spatial frequency of the grating along the x-axis, and $\phi$ is the initial phase of the grating at $x = 0$.

The definition of Fisher information is:

$$I(\theta) = E\left[\left(\frac{\partial L(\theta)}{\partial \theta}\right)\left(\frac{\partial L(\theta)}{\partial \theta}\right)^T\right], \tag{2}$$

where $\theta = [\alpha, k, \phi]^T$ are the parameters to be estimated and $L(\theta)$ is the log-likelihood function of $\theta$. $E[\cdot]$ is taking an expectation over all possible outcomes (possible noisy images). Assuming detected photons from the above object through a particular microscope system follow Poisson distribution due to photon counting noise, we calculate the Fisher information as (*27*, *28*)

$$I(\theta) = \sum_i \frac{1}{u_i}\left(\frac{\partial u_i}{\partial \theta}\right)\left(\frac{\partial u_i}{\partial \theta}\right)^T, \tag{3}$$

where $\boldsymbol{u} = [u_1, \ldots, u_i, \ldots, u_N]$ represents data obtained from $obj$ through different imaging modalities and $u_i$ represents the expected photons of individual pixels. For wide-field and confocal systems, $\boldsymbol{u}$ would be the ideal image (**Methods**). For conventional SIM, $\boldsymbol{u}$ represents 9 frames of ideal images with different structured illumination patterns (**Methods**). For ISM, $\boldsymbol{u}$ would be images of emission patterns at different scanning positions (**Methods**).

Variance lower bound on the phase estimation, as well as frequency and relative amplitude, can be calculated through Cramér–Rao lower bound (CRLB) (*26*, *27*) as the inverse of the Fisher information matrix $I(\theta)$,

$$Var(\hat{\theta}) \geq CRLB(\theta) = I^{-1}(\theta). \tag{3}$$

The inequality indicates the difference between two matrices is positive semidefinite.

As Fisher information $I(\theta)$ is additive, the same object imaged in a larger field of view (FOV) would result in larger Fisher information (smaller CRLB), assuming photon flux per area is constant (**Fig. S1**). Therefore, we limit our investigation throughout the work to a FOV of 10 µm×10 µm. To further simplify this problem, we use the square root of CRLB of phase-$\sigma_\phi$, the phase estimation precision limit as a scalar measure indicating the microscope's resolving power (higher $\sigma_\phi$ means worse resolving power).

To define a resolution limit, one needs a criterion of 'resolved.' For example, Rayleigh's criterion, $\frac{0.61\lambda}{NA}$, is defined as the distance of two points where their point spread function (PSF) first minima reach each other's center (*29*). Sparrow et al., defines two points resolvable when the mid-point of their joint intensity profile shows a minimum (*30*). In search of a noise-considered resolution criterion, we examined the visualization of noisy images of wide range frequencies, photon densities, and phase estimation precision limits $\sigma_\phi$ (**Fig. S2**). Through visual comparison, we decided to set the threshold of phase estimation uncertainty $\sigma_\phi$ at 0.04 *rad* as the resolving criterion. Subsequently, IbR is defined as the reciprocal of the highest frequency the imaging modality could resolve given the above criterion (colored dashed line in **Fig. 2a**).

We want to note that the definition of IbR and $\sigma_\phi$ criterion is for the ease of interpretation, while other criteria such as $\sigma_\alpha$, $\sigma_k$, or Fisher information per area can also be formed. Using Fisher information per area as resolving power measure will forgo the need to a fixed FOV when defining a resolution criterion (**Fig. S3**). However, in contrast to $\sigma_\phi$-which can be stated as the best achievable uncertainty of phase, Fisher information does not have a direct physical meaning. Moreover, the correlation between all three parameters would not be considered if using Fisher information scalar term. As for other parameters—relative amplitude $\alpha$ and frequency $k$, we consider the phase $\phi$ has a closer relationship to resolution. Besides, all three parameters share a similar trend in our simulation (**Fig. S4**).

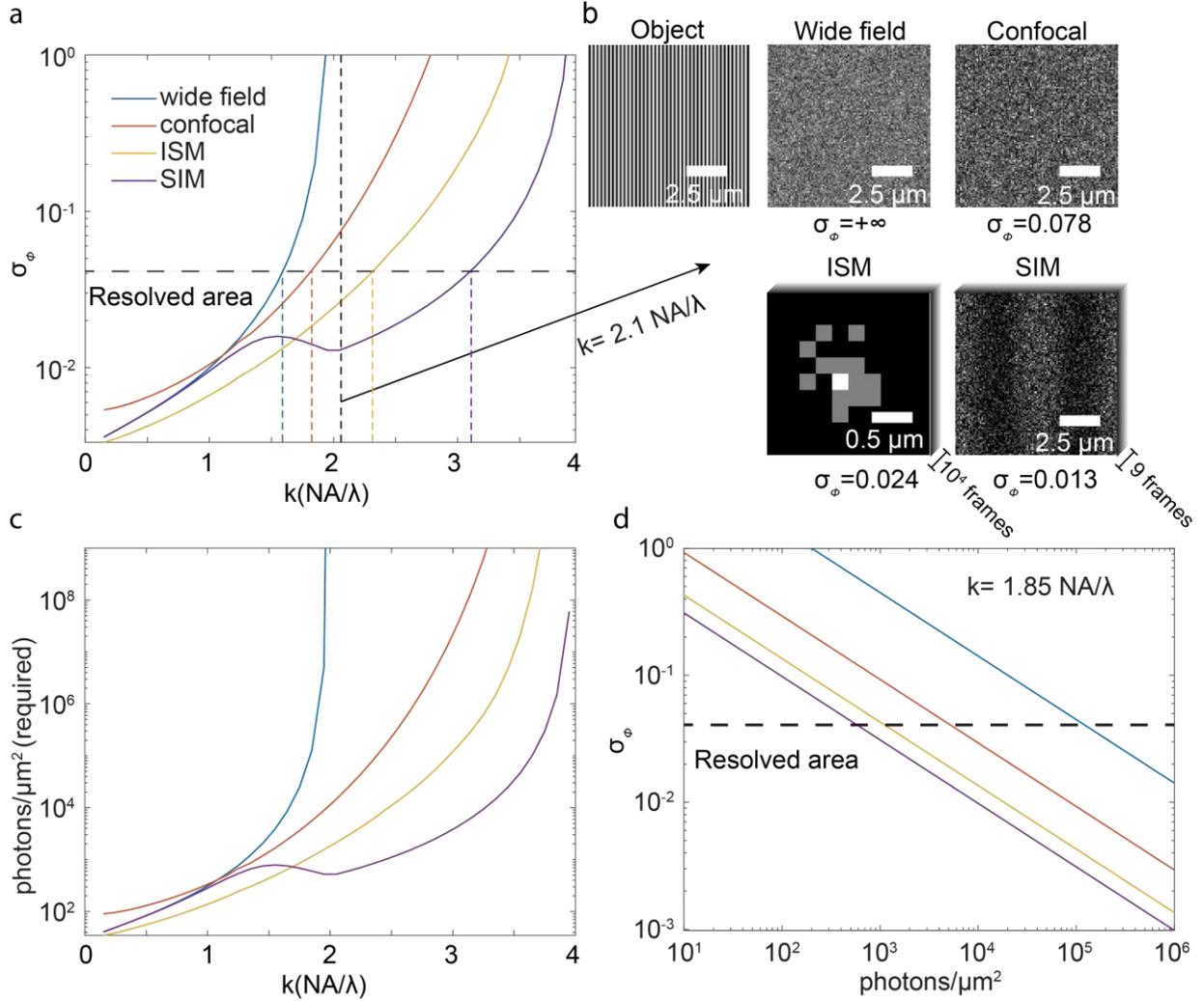

**Fig. 2.** Comparison of four imaging modalities, wide-field microscopy, confocal microscopy, SIM, and ISM at resolving different frequencies using phase estimation precision limit $\sigma_\phi$. (a) Resolving power in terms of $\sigma_\phi$ demonstrated for four imaging modalities. The lateral black dashed line ($\sigma_\phi = 0.04\ rad$) indicates the applied resolving criterion. Colored dashed lines indicate the reciprocal of IbR for each imaging modality. (b) Visualization of noisy images for an object of frequency $\frac{2.1NA}{\lambda}$, a frequency slightly higher than the diffraction limit. The raw image in the SIM is the image of shifted frequency of the original grating object, as the structured illumination frequency is at optical transfer function (OTF) boundary which would not appear in the image. (c) Photons per area required for different grating structures to reach the applied resolving criterion ($\sigma_\phi = 0.04\ rad$), ranging from 0 to 1 billion photons per μm². (d) Relationship between $\sigma_\phi$ and photon count for an object of frequency $\frac{1.85NA}{\lambda}$ plotted in log scale. The simulation conditions are set with photon emission density of 5000 photons/μm², numerical aperture of 1.4, immersion medium refractive index of 1.5, and emission wavelength of 0.7 μm. Confocal system considered a pinhole radius of 0.5 Airy disk Unit (AU). SIM implemented a structured illumination frequency of $k_{st} = \frac{2NA}{\lambda}$. Camera pixel size in the wide-field system and SIM as well as scanning intervals in the confocal system and ISM are set to 0.1 μm.

*Achievable resolving power in the presence of noise*

In IbR, photons are the information currency for resolving power. In extreme cases, when no photons are detected, images would not be formed, leaving no information (infinite $\sigma_\phi$) for all imaging modalities. In contrast, when photons are unlimited (noise negligible), IbR of various imaging modalities would reach their conventional (Abbe) resolution limits (**Fig. 2a, 2c**). With a limited photon (defined as the number of photons emitted per area, hereafter referred to as photon density), imaging modalities can no longer achieve their traditional limit due to Poisson noise. Instead, their practical resolving power differs drastically, and their differences also depend on the complexity of the imaging target (e.g., spatial frequency of the target). Using the developed approach, we investigate wide-field microscopy, confocal microscopy, SIM, and ISM, four commonly used imaging modalities, quantifying their resolving power in the presence of noise towards objects of frequencies ranging from 0 to $\frac{4NA}{\lambda}$.

The general trend of resolving power of the wide-field system, confocal system, SIM, and ISM targeting objects of different frequencies is demonstrated in Fig. 2a. At low frequencies, well within the diffraction limit, ISM achieves significantly superior resolving power surpassing all other modalities, whereas the wide-field system and SIM perform similarly. For example, within a low-frequency range ($[0, \frac{NA}{\lambda}]$), ISM has phase estimation precision limit $\sigma_\phi$ up to 1.5 times lower than that of the wide-field system and SIM (**Fig. 2a**). This can be explained by ISM's effective usage of small area pixel detectors which extends its OTF while the multi-pixel based PSF detection ensured no photon loss. In comparison, the confocal system has the worst resolving power at low frequencies among all modalities quantified due to photon loss by pinhole rejection. At frequency $\frac{0.2NA}{\lambda}$, a confocal system has a 1.5 times higher (worse) phase estimation precision limit $\sigma_\phi$ value than that of a wide-field system (**Fig. 2a**).

At higher frequencies, while the wide-field system reaches diffraction limit at frequency $\frac{2NA}{\lambda}$ resulting in an infinite $\sigma_\phi$, the confocal system, SIM and ISM demonstrate capacity overcoming the diffraction limit of the wide-field system (**Fig. 2a**). SIM showed the highest resolving power given a limited photon level. For a photon density of 5000 photons/μm² (**Fig. 2a**), SIM achieves an IbR of 170 nm, surpassing the conventional diffraction limit resolution 250 nm while ISM, confocal and wide-field systems achieved IbR of 220 nm, 270 nm and 315 nm, respectively (color dashed line in **Fig. 2a**). Such superiority results from the SIM-enabled shift of object's frequency component from outside the OTF boundary to the center of OTF which enhance its transfer magnitude (**Fig. S7, S8**).

The above results can also be demonstrated from the view of photon requirement. To achieve a specific practical resolution (IbR), imaging modalities would require different minimum numbers of photons (**Fig. 2c**). To resolve an object with a frequency as low as $\frac{0.2NA}{\lambda}$, a confocal system requires 90 photons/μm². In comparison, others require less than 40 photons/μm², a more than 2-fold difference. To resolve an object of frequency $\frac{NA}{\lambda}$, ISM requires 150 photons/μm² while other modalities require at least 380 photons/μm² emitted from the specimen. At a higher frequency $\frac{1.85NA}{\lambda}$ (slightly below the diffraction limit), a wide-field system requires 125,000 photons/μm². A confocal system requires 5580 photons/μm², roughly 20 times less than a wide-field system, while ISM and SIM require 2 orders fewer photons at 1147 photons/μm² and 600 photons/μm². To resolve an object of significantly higher frequency $\frac{3.5NA}{\lambda}$, SIM requires the least amount of photon density of 24,800 photons/μm² compared to 4,607,000 photons/μm² requirement for ISM, more than 2 orders of magnitude difference. In contrast, even with one billion photons/μm², a confocal system can't resolve such a structure, although the structure is theoretically resolvable by the confocal system based on its optical transfer function (*19, 31*).

The phase estimation precision limit $\sigma_\phi$ has an inverse square root relationship with photon counts (**Fig. 2d**). This is because as each incoming photon is considered as an event in Fisher information calculation, the Fisher information matrix (inverse of CRLB) is linearly proportional to photon counts in the absence of background fluorescence.

In microscopy imaging, the background is the unwanted photon signal which increases noise (*12*). To quantify background influence on resolving power, we separate the target object and the assumed uniform background by reformulating Eq. 1 as $obj(x) = U'_{ave}[1 + \sin(2\pi klx + \phi)] + bg$. Here $U'_{ave}$ is the photon density emitted by the target structure. $bg$ is the photon density from the background. At low frequencies, ISM is better tolerant of background photons while others perform similarly. Given a signal level $U'_{ave}$ 1000 photons/μm², to resolve an object of relatively low frequency as $\frac{0.95NA}{\lambda}$, wide-field system, confocal system, and SIM can tolerate background up to 5000 photons/μm², while ISM could tolerate background up to 12,000 photons/μm². At high frequencies, SIM and ISM are superior to the others. To resolve an object of relatively high frequency as $\frac{1.85NA}{\lambda}$, a frequency close to the diffraction limit, SIM can tolerate background of up to 800 photons/μm² while ISM could only resolve the structure when the background is close to zero. On the other hand, given the limited photon level, the wide-field system and the confocal system cannot resolve the structure even in a background-free environment (**Fig. S5**).

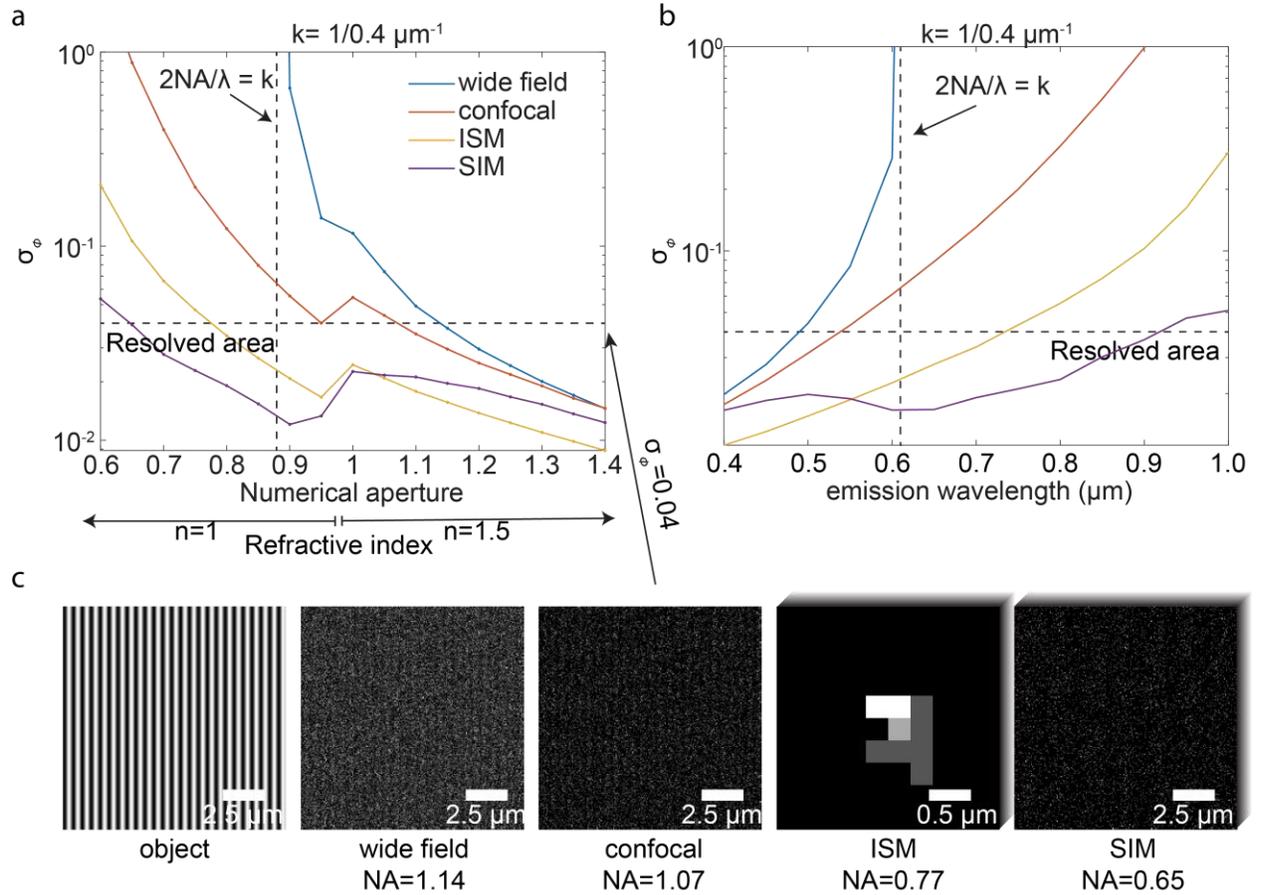

**Fig. 3.** Influence of numerical aperture and emission wavelength on phase estimation precision limit $\sigma_\phi$. (a) Influence of numerical aperture on $\sigma_\phi$. The emission wavelength is set to 700 nm. (b) Influence of wavelength on $\sigma_\phi$. Numerical aperture is set to 0.8 and refractive index is set to 1. (c) Visualizations of the black dashed line $\sigma_\phi = 0.04$ in various imaging modalities. In the above plots, we set the photon emission density as 5000 photons/μm². The confocal system chooses the pinhole radius 0.5 AU. SIM has the structured illumination frequency $k_{st} = \frac{2NA}{\lambda}$. Camera pixel size in the wide-field system and SIM as well as scanning intervals in the confocal system and ISM are set to 0.1 μm.

*NA and emission wavelength influence on resolving power*

While in Abbe's resolution limit ($\frac{\lambda}{2NA}$), numerical aperture (NA) and emission wavelengths share an equally important role. Their influence on IbR (noise considered resolution), however, differs. In addition to the effect on OTF, increasing NA also increases the photon collection angle, thus making NA a more critical factor than the wavelength in IbR. For an object of frequency $\frac{1}{0.4}$ $\mu m^{-1}$, quantified through $\sigma_\phi$ in ISM, the precision limit improved 5.96 times when increasing NA from 0.6 to 0.8, whereas an equivalent change in detection wavelength from 0.8 to 0.6 μm only improves it by 2.45 times (**Fig. 3a, 3b**).

Given a specific photon density, IbR also enables us to calculate the minimum requirement of NA and emission wavelength for various imaging modalities to resolve an object. As an example, we examine a grating object with a frequency of $\frac{1}{0.4}$ $\mu m^{-1}$ (peak to peak interval of the grating object is 0.4 μm) with photon density 5000 photons/μm² as an instance. When the emission wavelength is fixed as 700 nm, it is theoretically resolvable with NA greater than 0.9 in a wide-field system without noise. However, when considering noise, a wide-field system requires an NA of 1.15 and thus requires an oil/water immersion objective. In contrast, confocal system, ISM, and SIM have much relaxed requirement with a minimum NA of 0.95, 0.75, and 0.65, respectively (**Fig. 3a**). When the NA of the microscope is fixed (e.g., 0.8 in air), wide-field system, confocal system, ISM, and SIM require emission wavelengths smaller or equal to 480 nm, 530 nm, 720 nm, and 900 nm to resolve the structure (**Fig. 3b**).

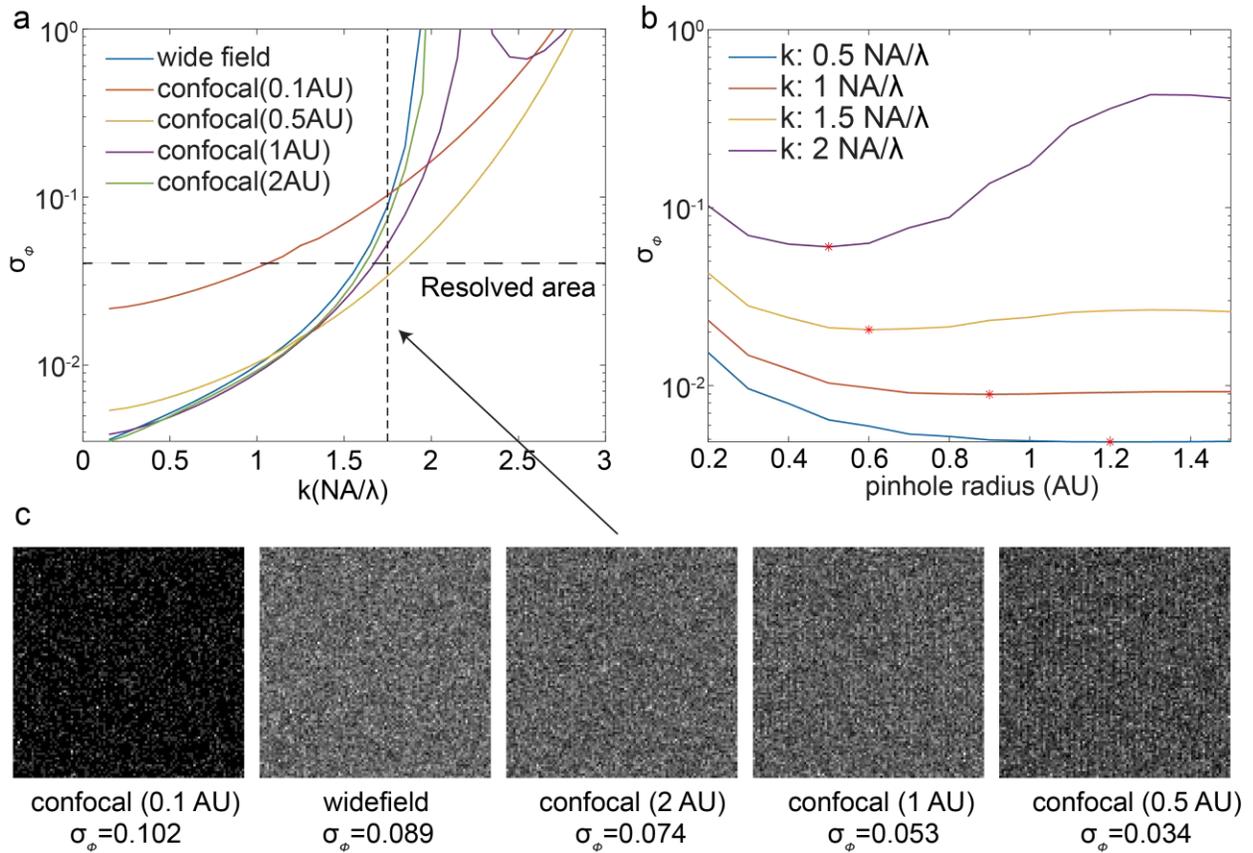

**Fig. 4.** Phase estimation precision limit $\sigma_\phi$ in a confocal system with different circular pinhole radii. (a) $\sigma_\phi$ comparison among a wide-field system and confocal systems with different pinhole radii with respect to target objects' frequencies. (b) For objects of a specific frequency, the relationship between pinhole radius and $\sigma_\phi$. The red star indicates where $\sigma_\phi$ reaches the minimum. (c) Visualization of noisy images of the object of frequency $k = 1.75 \frac{NA}{\lambda}$. In the above plots, we set the photon emission density as 5000 photons/μm², numerical aperture as 1.4, immersion medium refractive index as 1.5 and emission wavelength as 0.7 μm. Camera pixel size in the wide-field system and scanning intervals in the confocal systems are set to 0.1 μm.

*Pinhole size influence on confocal system*

In confocal imaging, shrinking pinhole size affects the resolving power in two opposite ways—improving it by broadening effective OTF while worsening it by decreasing photon detection due to photon rejections of the pinhole (*19*, *24*) (**Supplement Note**, **Fig. S6**). When the pinhole radius is too large (e.g., 2 Airy disk units (AU)), the confocal system's performance would be close to that of a wide-field system (**Fig. 4a**) (*6*,*7*). As the pinhole radius shrinks, the confocal system tends to have better resolving power. For instance, confocal systems with pinhole radii of 1 AU ($AU = \frac{1.22\lambda}{NA}$) and 0.5 AU result in IbR of 300 nm and 270 nm compared to that of the wide-field system at 315 nm (**Fig. 4a**). However, when pinhole size shrinks even smaller, the resolving power of confocal system decay rapidly due to photon rejection of pinhole. For example, a confocal system with a pinhole radius of 0.1 AU results in an IbR of 500 nm. The tradeoff between improvement and deterioration of the confocal system's resolving power, often balanced by pinhole size from experience, can now be quantified through phase estimation precision limit $\sigma_\phi$ to find the optimal pinhole radius. We found over a large range frequency, 0.5 AU radius of pinhole size reaches the lowest $\sigma_\phi$ in general, suggesting that it is a proper setting for confocal systems in most cases which is in agreement with the common practice in confocal systems (*24*). To seek an optimal resolving power for an object at specific frequencies, Fig. 4b demonstrates the optimal pinhole radius to achieve the lowest $\sigma_\phi$ given four grating objects of different frequencies. For objects of frequency $\frac{0.5NA}{\lambda}, \frac{NA}{\lambda}, \frac{1.5NA}{\lambda}$ and $\frac{2NA}{\lambda}$, the optimal pinhole radius is 1.2, 0.9, 0.6, 0.5 AU, respectively. Generally, a small pinhole radius suits objects of high frequencies, while a large pinhole radius suits objects of low frequencies.

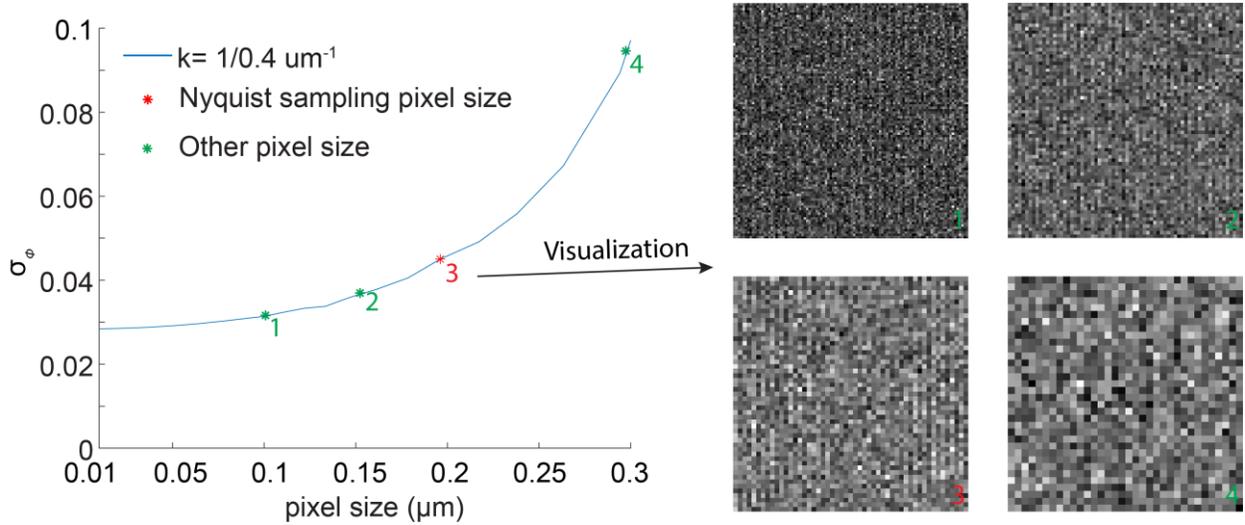

**Fig. 5.** Camera pixel size influence on resolving power of wide-field microscopy in terms of $\sigma_\phi$ (left). Pixel size is denoted as the pixel size mapped on the object plane. Visualization of noisy images taken with different pixel sizes given a fixed emitted photon count (right). Imaging conditions are set with a photon emission density of 5000 photons/μm², numerical aperture of 1.4, immersion medium refractive index of 1.5, and emission wavelength of 0.7 μm.

*Pixel size influence on the wide-field microscope*

The pixel size of the digital image detector, often considered irrelevant to the conventional resolution limit, however, has an impact on IbR. Generally, while ignoring the camera readout noise (instrument noise), a smaller pixel size provides greater resolving power (better IbR). We show this trend even when pixel size gets smaller than that required by the Nyquist sampling theorem (*32*, *33*). Taking wide-field microscopy as an example, the phase estimation precision limit $\sigma_\phi$ increases twice from 0.05 rad to 0.1 rad while changing pixel size from 0.2 μm (Nyquist sampling pixel size) to 0.28 μm, indicating the importance of using pixel size smaller than Nyquist sampling requirement (**Fig. 5**). In addition, we investigated IbR in situations of applying a pixel size even smaller than that required by Nyquist sampling. We found that further reducing pixel size to 0.1 μm would decrease $\sigma_\phi$ by roughly 30% compared to a Nyquist sampling pixel size of 0.2 μm, approaching its limit of 0.029 rad when using infinite small pixel size (approximated as 0.01 μm). The above results indicate that binning pixels as equivalent to using larger pixel sizes in a microscope will reduce the effective resolving power of the imaging system when instrument noise is negligible. This is understood as each pixel being a photon bucket that integrates all photons falling into its active area, the pixelization process function as a low pass filter (*34*) (**Supplement Note**).

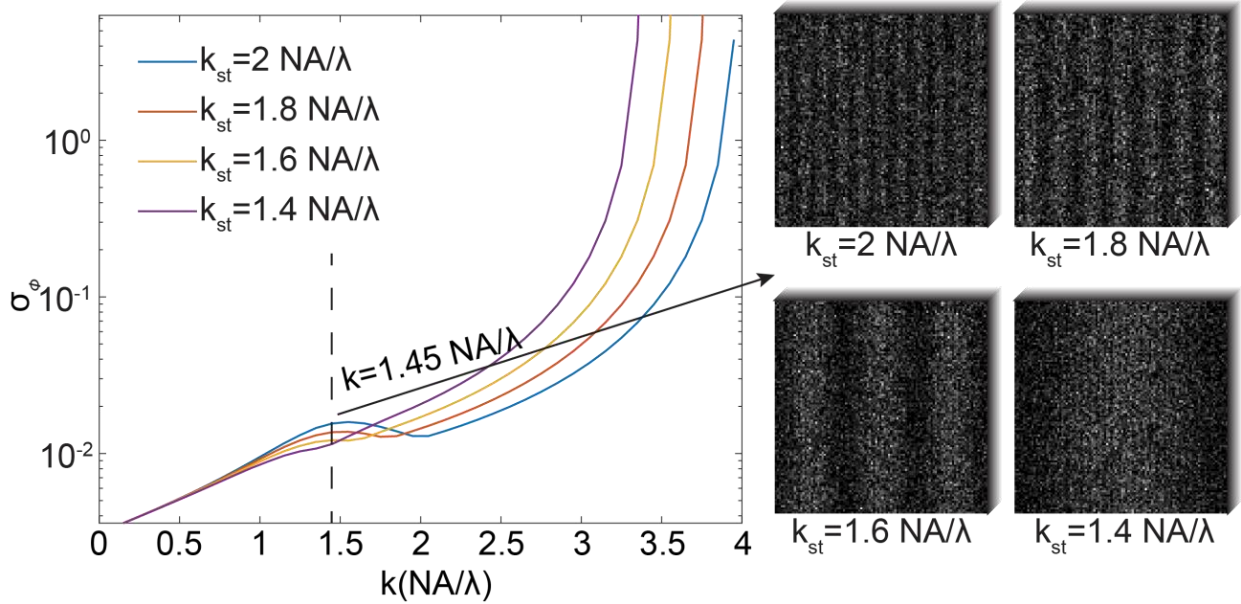

**Fig. 6** Influence of the structured illumination frequency on SIM performance. $\sigma_\phi$ of SIM with four structured illumination frequencies are quantified for objects at different frequencies (left). The visualizations of noisy images

of SIM with different structured illumination frequencies given an object of frequency $\frac{1.45NA}{\lambda}$ are shown on the right. In the above plots, we set the photon emission density as 5000 photons/μm², numerical aperture as 1.4, immersion medium refractive index as 1.5, and emission wavelength as 0.7 μm. The camera pixel size in SIM is set to 0.1 μm.

*Influence of the structured illumination frequency on SIM*

In aberration-free systems, it is a common intuition that complex objects (objects with high-frequency components) are harder to resolve. In terms of IbR, imaging modalities generally achieve greater resolving power (lower $\sigma_\phi$) at lower frequencies than that at higher frequencies (Fig. 2a). However, there are exceptions while using SIM in some frequency ranges that a higher frequency structure is better resolved than a lower frequency one. Taking SIM with an illumination frequency $k_{st} = \frac{2NA}{\lambda}$ as an example (blue curve in **Fig. 6**), phase estimation precision limit $\sigma_\phi$ at frequency $\frac{2NA}{\lambda}$ is smaller (0.013 rad) than that at frequency $\frac{1.5NA}{\lambda}$ (0.016 rad, **Fig. 6**). Since the emission from structured illuminated object contains information from both its original and shifted spatial frequencies, with increasing the object's frequency, the transfer amplitude of its original frequency would decrease as it approaches the boundary of OTF. Meanwhile, the transfer amplitude of its shifted frequency would increase as it approaches the center of OTF (**Fig. S7, S8**). As both original and shifted frequency contributes to the phase estimation, the opposite trend of their resulting transfer amplitudes causes variations in the relationship between phase estimation precision limit $\sigma_\phi$ and object's frequency $k$.

As the frequency composition of a structured illuminated object is influenced by its structured illumination frequency-$k_{st}$, we ask how SIM of different structured illumination would influence its resolving power. We tested SIM with four illumination frequencies. The general trend suggests a higher illumination frequency-$k_{st}$ provides lower phase estimation precision $\sigma_\phi$ (better resolving power) over the most range of object's frequencies (**Fig. 6**). This agrees that the commonly used illumination frequency in SIM is at the OTF boundary (*20, 21*). However, in some cases, a lower illumination frequency could improve the resolving power compared to that of a higher frequency, contradicting this common practice. For example, while imaging an object of frequency $\frac{1.5NA}{\lambda}$, $\sigma_\phi$ of SIM is worse when using an illumination frequency of $k_{st} = \frac{2NA}{\lambda}$, a frequency at the OTF boundary, than that of a lower illumination frequency (e.g., $k_{st} = \frac{1.4NA}{\lambda}$, **Fig. 6a**). Above results indicate that in theory using higher structured illumination frequency in SIM is generally beneficial but not always.

In conventional SIM, the structured illumination pattern is a 1D sinusoidal wave pattern that only shifts the object's frequency along one direction. The common practice is to apply a set of three patterns, each having 60-degrees orientation difference, to cover the 2D frequency plane, which causes an uneven resolution distribution along different directions (*20, 21*). Quantifying through our phase estimation precision limit $\sigma_\phi$, we measured this orientation-dependent resolution in **Fig. S9**. In the condition of photon density 5000 photons/µm², 30 degrees difference in the initial orientations of structured illumination patterns set could alter IbR of SIM from 150 nm to 192 nm along one direction. Increasing the number of illumination pattern orientations in a set could alleviate the uneven distribution. In the above conditions, we tested that applying a set of six illumination patterns instead of three in SIM achieved an IbR of 170 nm for all initial orientations (**Fig. S9**).

**Discussion**

The purpose of IbR is to create a noise-considered resolution limit predicting the performance of imaging modalities with finite photon counts. The concept of IbR relies on the ideal image formation of a periodic object. In our calculation, we assumed the fluorescence response is linear, meaning the emission intensity is proportional to the illumination power. Thus, our IbR cannot be directly applied to some of the super-resolution imaging modalities, such as single-molecule localization microscopy (SMLM) (*35–38*) and stimulated emission depletion (STED) microscopy (*39, 40*), which rely on nonlinear fluorescence response. For example, SMLM requires the 'blinking' of individual emitters. However, the emitters switch stochastically, and thus, the resolution limit of SMLM relies on the exact on-off time sequences of imaged single molecules, which is challenging to summarize for IbR. The resolution of STED depends on the power of the depletion laser and its PSF. In an ideal situation where the depletion PSF has a perfect donut shape and infinite power, the resolution of STED can reach the molecule's size (*40*). IbR could potentially provide a method for assessing its practical resolution when provided the properties of the non-linear behavior of the probe and its physical model during depletion. In addition, another limitation of IbR is that it only quantifies the lateral resolution in a 2D sample at the current stage. Without considering the actual 3D environment in the sample, IbR cannot measure the ability to resist out-of-focus background, for which the confocal system is designed.

Fisher information allowed us to consider photon noise in our resolution measure, and the phase estimation precision limit $\sigma_\phi$ can be understood as the uncertainty of lateral positioning of the grating pattern. Although CRLB does not guarantee any estimators approaching it, it is often considered the best achievable precision of a parameter. This is because the variance of an unbiased maximum likelihood estimator (MLE) asymptotically approaches CRLB (*26*).

In modern microscopy methods, the raw data are often post-processed to form the image for visualization. A question naturally arises as to whether data post-processing could increase the information and, therefore resolution (e.g., IbR). To this end, our theoretical derivation (**Supplement Note**) shows that no Fisher information could be increased through deterministic data post-processing methods. Therefore, post-processing methods including image reconstruction algorithms, could either keep IbR constant or worsen IbR.

IbR provides a new measure of quantifying the practical resolving power of microscopy imaging modalities considering finite photons. The noise-considered resolution measure could provide a theoretical and statistical reference for fluorescence microscope imaging modalities in photon-limited conditions and also serve as a guide for the further development of novel microscopy methods.

**Methods**

*2D periodic structure generation and microscope imaging system simulation*

We simulate a 2D single-tone sinusoidal wave grating object by Eq. 1 in a 20 μm×20 μm region, which contains 4000×4000 pixels (5 nm per pixel). Throughout the simulation except for when background level is specifically examined, the relative amplitude $\alpha$ is set to 0.95 (95% signal, 5% background), and the initial phase $\phi$ is set to 0. The ideal image is simulated as projected on the object plane, so the magnification is unitary. We simulate the ideal image formation process based on Fresnel approximation (*4*) without optical aberrations. This process can be viewed as object convolution with PSF (*4*). To generate PSF, we first create a pupil function of circular shape with radius $\frac{NA}{\lambda}$ in the Fourier space. In Fourier space, each pixel denotes frequency interval $\frac{1}{L}$ where $L = 20 \ \mu m$ is the length of the image. The radius of the pupil function in terms of the number of pixels is $round(\frac{NA}{\lambda} * L)$, where $round$ function make $\frac{NA}{\lambda} * L$ an integer. PSF is the magnitude square of the Fourier transform of the pupil function. The ideal image simulation conditions for each figure are listed in the figure caption, including the numerical aperture of the objective, refractive index of the immersion media, fluorescence emission wavelength, and expected photon density.

*Wide-field microscopy simulation*

In wide-field microscopy simulation, we directly convolve the object with the PSF to obtain the ideal image. After convolution, to avoid the simulation error at the boundary, we crop the central region 10 μm×10 μm from the ideal image to form the FOV. The cropped image is subsequently binned to 100 pixels ×100 pixels to mimic the pixel integration effect of camera detection, except for Fig. 5, where pixel size in the wide-field system is specifically examined. In the 100 pixels ×100 pixels image, the camera pixel size is 0.1 μm. The binned image representing $\boldsymbol{u}$ in Eq. 1, is used to further calculate the Fisher information matrix.

*Confocal microscopy simulation*

In confocal system simulation, we assume the excitation laser focus is diffraction limited, sharing the same numerical aperture with the emission detection. Although the emission wavelength is typically larger than the excitation wavelength due to Stokes shift of the fluorescence emission, for simplicity of mathematical calculation, we assumed the excitation shares the same wavelength as emission. This assumption allows us to waive one additional parameter, simplifying the simulation model. As a result, the excitation PSF is identical to the emission PSF throughout our simulation. At the image plane, we simulate a circular pinhole in the conjugate plane with the center of the excitation PSF. The radius of the pinhole we simulated in each figure are listed in the figure captions. The image recording sensor is simulated as a bucket photon detector which integrates the photons collected for each scanning position. For each position, we dot product the object with the excitation PSF to generate the excited object fluorescence distribution. Subsequently, we convolve this excited object with PSF to form the image before the pinhole. We dot product the image with the pinhole mask (1 within the pinhole radius, 0 otherwise), sum the result, and assign it to the scanning position. Throughout the simulation, we set the total number of scanning positions to 100×100 covering the central 10 μm×10 μm region with a scanning interval of 0.1 μm. The image recorded with 100*100 scanning points representing $\boldsymbol{u}$ in Eq. 1 is used to further calculate the Fisher information matrix.

*Image scanning microscopy simulation*

ISM simulation follows the same steps as the confocal system simulation except using an array detector instead of a bucket detector with a physical pinhole. We record the image at each scanning position with a camera of 10 pixels×10 pixels with a pixel size of 0.2 μm, whereas the maximum pixel size according to Nyquist sampling is 0.25 μm. We set the total number of scanning positions to be 100×100 covering the central 10 μm×10 μm region with a scanning interval of 0.1 μm. Therefore, in total, $10^4$ frames of images of different scanning positions formed the data $\boldsymbol{u}$ in Eq. 1, which is used to further calculate Fisher information.

*Structured illumination microscopy simulation*

In SIM, to form the ideal image, we first applied the structured illumination pattern on the object to form a structurally-illuminated object. Throughout the work, we used the structured illumination pattern of a single-tone sinusoidal wave. The oscillation range of the structured illumination intensity is from 0 to 2, such that applying the pattern to the object would not affect the total number of photons the object emits. The frequencies of the structured illumination pattern simulated in each figure are listed in the figure captions. We convolve the structurally-illuminated object with PSF to form the ideal image. Like wide-field microscopy, we crop the central region 10 μm×10 μm of the ideal image, and then the cropped image is binned to 100 pixels ×100 pixels to mimic the actual camera detection, where the camera pixel size is 0.1 μm. In total, we collect images with three structured illumination orientations of 60º difference, and each orientation contains three patterns with a phases difference of $\frac{2\pi}{3}$. A total of nine image frames with different structure illumination orientations and phases are collected as data $\boldsymbol{u}$ in Eq. 1, which is used to further calculate Fisher information.

*Fisher information calculation*

When a random variable (in this case: pixel value) follows Poisson distribution, the Fisher information can be calculated by Eq. 3, which requires the derivative of $\boldsymbol{u}$ with respect to $\boldsymbol{\theta} = [\alpha, k, \phi]$. We calculate it numerically by the difference of the $\boldsymbol{u}$ function with a small increment and decrement on the specific parameter. For example, we calculate $\frac{\partial u_i}{\partial \alpha} = \frac{u_i(\alpha+\Delta\alpha,k,\phi) - u_i(\alpha-\Delta\alpha,k,\phi)}{2\Delta\alpha}$ where we set $\Delta\alpha$ equal to 0.01 in our simulation where $\alpha$ is the modulation amplitude ($\alpha \in [0,1]$) of the periodic object. Respectively, we use $\Delta\phi = 0.01$ and $\Delta k = \frac{1}{L}$ where $L = 20\ \mu m$ is the total length of the image, which are both smaller than 2% of our typical simulations' range. By arranging $\Delta k$ this way, we will ensure that in the frequency domain, the Fourier transform of a single-tone sinusoidal pattern provides isolated frequency peaks in single pixels.

**Acknowledgments:** We thank Sheng Liu for her discussions on this topic, Vamara Dembele, Maryam Mahmoodi, Peiyi Zhang and Hao-Cheng Gao for suggestions during the writing of the manuscript.

**Funding:** National Institute of General Medical Sciences (R35GM119785).

**Competing interests:** The authors declare that they have no competing interests.

**Data and materials availability:** Codes needed to evaluate the conclusions in the paper are present in github. Additional data related to this paper may be requested from the authors.

`

# Supplementary Materials for

A statistical resolution measure of fluorescence microscopy with finite photons

Yilun Li[1], Fang Huang[1,2,3]∗

Correspondence to: fanghuang@purdue.edu

**This PDF file includes:**

Supplementary Text: Note1-Note3

Figs. S1-S10

**Supplementary Text**

Note 1: Data post-processing influence on Fisher information

Supposing we have 2 random variables which are $J$ and $K$ denoting raw data and processed data. $K$ is determined by $J$ by function $K = H(J)$ where $H$ is a deterministic function. Consider the chain rule of Fisher information (*41*)

$$I_{J,K}(\theta) = I_J(\theta) + I_{K|J}(\theta).$$

The conditional Fisher information is

$$I_{K|J}(\theta) = E_j[I_{K|J=j}(\theta)].$$

By Fisher information definition we have

$$I_{K|J=j}(\theta) = E_k\left[\left(\frac{\partial \log(p(K|J=j;\theta))}{\partial \theta}\right)^2\right],$$

where $p(K|J = j; \theta)$ is the conditional probability density function. Since $K = H(J)$ is determined as

$$p(K|J = j; \theta) = \begin{cases} 1 & K = H(j) \\ 0 & otherwise \end{cases},$$

above equation indicates the conditional probability density function is determined on condition but irrelevant to any parameter $\theta$. Thus, the partial derivative is zero resulting in

$$I_{K|J=j}(\theta) = 0, \forall j,$$
$$I_{K|J}(\theta) = E_j[I_{K|J=j}(\theta)] = 0.$$

Combining first equation we have

$$I_J(\theta) = I_{J,K}(\theta) = I_K(\theta) + I_{J|K}(\theta) \geq I_K(\theta).$$

Last step of inequality comes from fact that any Fisher information is greater or equal to 0 (*26*). If the processing is deterministic, then Fisher information in the processed data $K$ would always be smaller or equal to raw data $J$. In another word, deterministic image post-processing cannot increase Fisher information of any parameter. Note that this statement is under condition that image processing is deterministic. If other prior information could be accessed, this relationship would not stand.

Note 2: Pixel binning effect

Pixel binning effect is stated in work (*33*). However, to the best knowledge of the author, I did not find a clear derivation on the pixel binning effect. We derive this effect as follows.

Supposing we have a 2D image with coordinate $(x, y)$, the image is recorded with discrete pixel index $[n, m]$. The function of each individual pixel is to integrate all photons falls within the pixel

$$Img[n,m] = \int_{x_n-\frac{d}{2}}^{x_n+\frac{d}{2}} \int_{y_m-\frac{d}{2}}^{y_m+\frac{d}{2}} Img(x,y) dx dy.$$

Here $(x_n, y_m)$ is the coordinate which is at the center of pixel $[n, m]$ and $d$ is the pixel size. $Img(x, y)$ is photon distribution function. $Img[n, m]$ is the image value at each pixel. We create a function defined as

$$f(x,y) = \frac{1}{d^2} \int_{x-\frac{d}{2}}^{x+\frac{d}{2}} \int_{y-\frac{d}{2}}^{y+\frac{d}{2}} Img(x',y') dx' dy'.$$

Then

$$Img[n,m] = f(x_n, y_m) d^2.$$

Applying Fourier transform on function $f(x, y)$ we have

$$\hat{f}(\xi, \eta) = \frac{1}{d^2} \iint_{-\infty}^{+\infty} f(x,y) e^{-2\pi i x \xi} e^{-2\pi i y \eta} dx dy$$

$$= \frac{1}{d^2} \iint_{-\infty}^{+\infty} [\int_{x-\frac{d}{2}}^{x+\frac{d}{2}} \int_{y-\frac{d}{2}}^{y+\frac{d}{2}} Img(x',y') dx' dy'] e^{-2\pi i (y\eta + x\xi)} dx dy.$$

Let $x' = x + \delta x$, $y' = y + \delta y$, we have

$$= \frac{1}{d^2} \iint_{-\infty}^{+\infty} [\int_{\frac{d}{2}}^{\frac{d}{2}} \int_{\frac{d}{2}}^{\frac{d}{2}} Img(x + \delta x, y + \delta y) d\delta x d\delta y] e^{-2\pi i (y\eta + x\xi)} dx dy.$$

Exchange the integration sequence we have

$$= \frac{1}{d^2} \int_{\frac{d}{2}}^{\frac{d}{2}} \int_{\frac{d}{2}}^{\frac{d}{2}} [\iint_{-\infty}^{+\infty} Img(x + \delta x, y + \delta y) e^{-2\pi i (y\eta + x\xi)} dx dy] d\delta x d\delta y$$

$$= \frac{1}{d^2} \int_{\frac{d}{2}}^{\frac{d}{2}} \int_{\frac{d}{2}}^{\frac{d}{2}} \widehat{Img}(\xi, \eta) e^{2\pi i (\delta y \eta + \delta x \xi)} d\delta x d\delta y$$

$$= \widehat{Img}(\xi, \eta) \frac{\sin(\pi \xi d)}{\pi \xi d} \frac{\sin(\pi \eta d)}{\pi \eta d}.$$

From expression, function $f(x, y)$ is function $Img(x, y)$ frequency filtered by two sinc functions. The image pixels' value $Img[n, m]$ is the sampling on function $f(x_n, y_m)$. Pixel size $d$ not only affects the sampling process but also affects the frequency filter applied on the image.

Note 3: Effective OTF of confocal

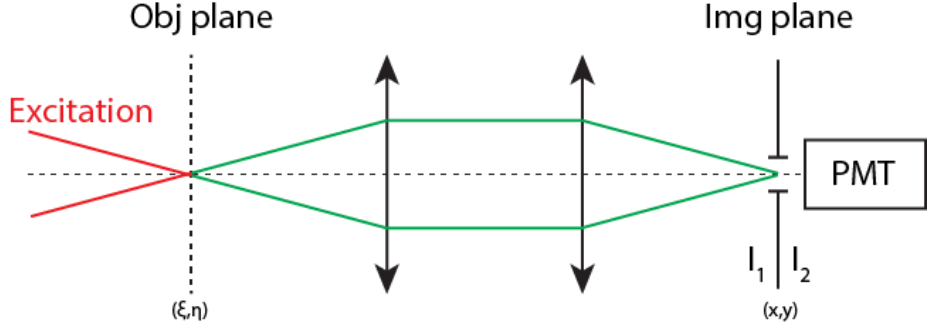

**Diagram of a confocal microscope.**

The OTF derivation of confocal can be found in works (*19,24*). For completeness and reference, we provide our version of derivation below.

The fluorescence distribution on the object plane is

$$I_{obj}(\xi,\eta) = obj(\xi,\eta) \cdot PSF_{ex}(\xi - \xi_0, \eta - \eta_0) \cdot \kappa,$$

where $obj(\xi,\eta)$ is the fluorophore density distribution, $PSF_{ex}(\xi,\eta)$ is the excitation PSF, $(\xi_0, \eta_0)$ denotes the scanning position and $\kappa$ is the excitation to emission coefficient we assumed to be 1 for simplicity. We also assumed the magnification is 1 for simplicity.

Forming the ideal image before pinhole we have

$$I_1(x,y) = \iint_{-\infty}^{+\infty} PSF(x - \xi, y - \eta) I_{obj}(\xi, \eta) d\xi d\eta.$$

The ideal image after pinhole would be

$$I_2(x,y) = I_1(x,y) \cdot I_{ph}(x - x_0, y - y_0),$$

where $I_{ph}(x,y)$ is the pinhole function either zero or one, $(x_0, y_0)$ is the conjugate position of scanning position on the image plane. Since magnification is 1, the conjugate relationship would be $(x_0, y_0) = (-\xi_0, -\eta_0)$.

In confocal imaging, a bucket photon detector would collect all photons for each scanning position and form a value. Therefore, the ideal image value of confocal at $(x_0, y_0)$ would be

$$I_{img}(x_0, y_0) = \iint_{-\infty}^{+\infty} I_2(x,y) dx dy$$

$$= \iint_{-\infty}^{+\infty} I_1(x,y) \cdot I_{ph}(x - x_0, y - y_0) dx dy$$

$$= \iint_{-\infty}^{+\infty} [\iint_{-\infty}^{+\infty} PSF(x - \xi, y - \eta) I_{obj}(\xi, \eta) d\xi d\eta] \cdot I_{ph}(x - x_0, y - y_0) dx dy.$$

Exchange the integration order we have

$$= \iint_{-\infty}^{+\infty} [\iint_{-\infty}^{+\infty} PSF(x - \xi, y - \eta) I_{ph}(x - x_0, y - y_0) dx dy] \cdot I_{obj}(\xi, \eta) d\xi d\eta.$$

If we assume the pinhole function is symmetric, we have

$$= \iint_{-\infty}^{+\infty} [\iint_{-\infty}^{+\infty} PSF(x-\xi, y-\eta) I_{ph}(x_0-x, y_0-y) dx dy] \cdot I_{obj}(\xi,\eta) \, d\xi d\eta.$$

Let $x' = x - \xi$ and $y' = y - \eta$, the equation becomes

$$= \iint_{-\infty}^{+\infty} [\iint_{-\infty}^{+\infty} PSF(x', y') I_{ph}(x_0+\xi-x', y_0+\eta-y') dx' dy'] \cdot I_{obj}(\xi,\eta) \, d\xi d\eta.$$

The integration in the bracket above is considered as a convolution between $PSF$ and $I_{ph}$ taking the value at $(x_0 + \xi, y_0 + \eta)$, then the equation becomes

$$= \iint_{-\infty}^{+\infty} (PSF \otimes I_{ph})(x_0+\xi, y_0+\eta) \cdot I_{obj}(\xi,\eta) \, d\xi d\eta.$$

Consider the conjugate relationship $(x_0, y_0) = (-\xi_0, -\eta_0)$ and we have

$$= \iint_{-\infty}^{+\infty} (PSF \otimes I_{ph})(\xi-\xi_0, \eta-\eta_0) \cdot I_{obj}(\xi,\eta) \, d\xi d\eta.$$

Expand the fluorescence distribution expression we have

$$= \iint_{-\infty}^{+\infty} (PSF \otimes I_{ph})(\xi-\xi_0, \eta-\eta_0) \cdot PSF_{ex}(\xi-\xi_0, \eta-\eta_0) \cdot obj(\xi,\eta) \, d\xi d\eta$$

$$= \iint_{-\infty}^{+\infty} (PSF \otimes I_{ph} \cdot PSF_{ex})(\xi-\xi_0, \eta-\eta_0) \cdot obj(\xi,\eta) \, d\xi d\eta.$$

If we consider $PSF$ and $PSF_{ex}$ are also symmetric, the equation becomes

$$= \iint_{-\infty}^{+\infty} (PSF \otimes I_{ph} \cdot PSF_{ex})(\xi_0-\xi, \eta_0-\eta) \cdot obj(\xi,\eta) \, d\xi d\eta.$$

Writing in this expression allows us to identify the ideal image formation in confocal as object convolution with an effective PSF where

$$PSF_{eff} = PSF \otimes I_{ph} \cdot PSF_{ex}.$$

The corresponding effective OTF would be Fourier transform on the PSF expression

$$OTF_{eff} = OTF \cdot FT\{I_{ph}\} \otimes FT\{PSF_{ex}\}.$$

Figures of effective OTF of confocal with different pinhole radii are plotted in Fig. S6.

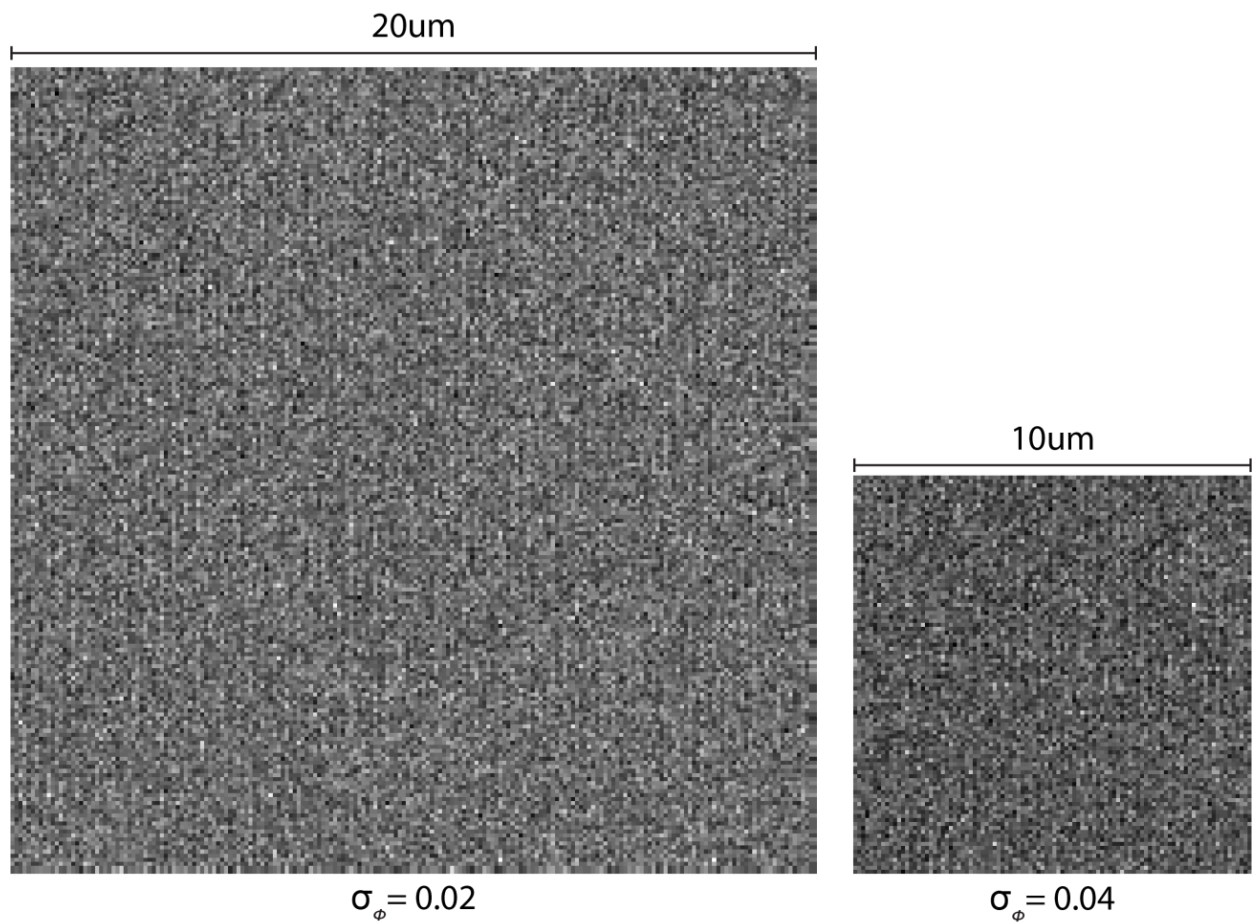

**Fig. S1.** $\sigma_\phi$ calculation of wide-field system on different FOV with constant photon density of 5000 photons/$\mu m^2$. Images are simulated under the condition: NA=1.4, emission wavelength $\lambda$=0.7 µm, immersion media refractive index n=1.5, and camera pixel size 0.1 µm.

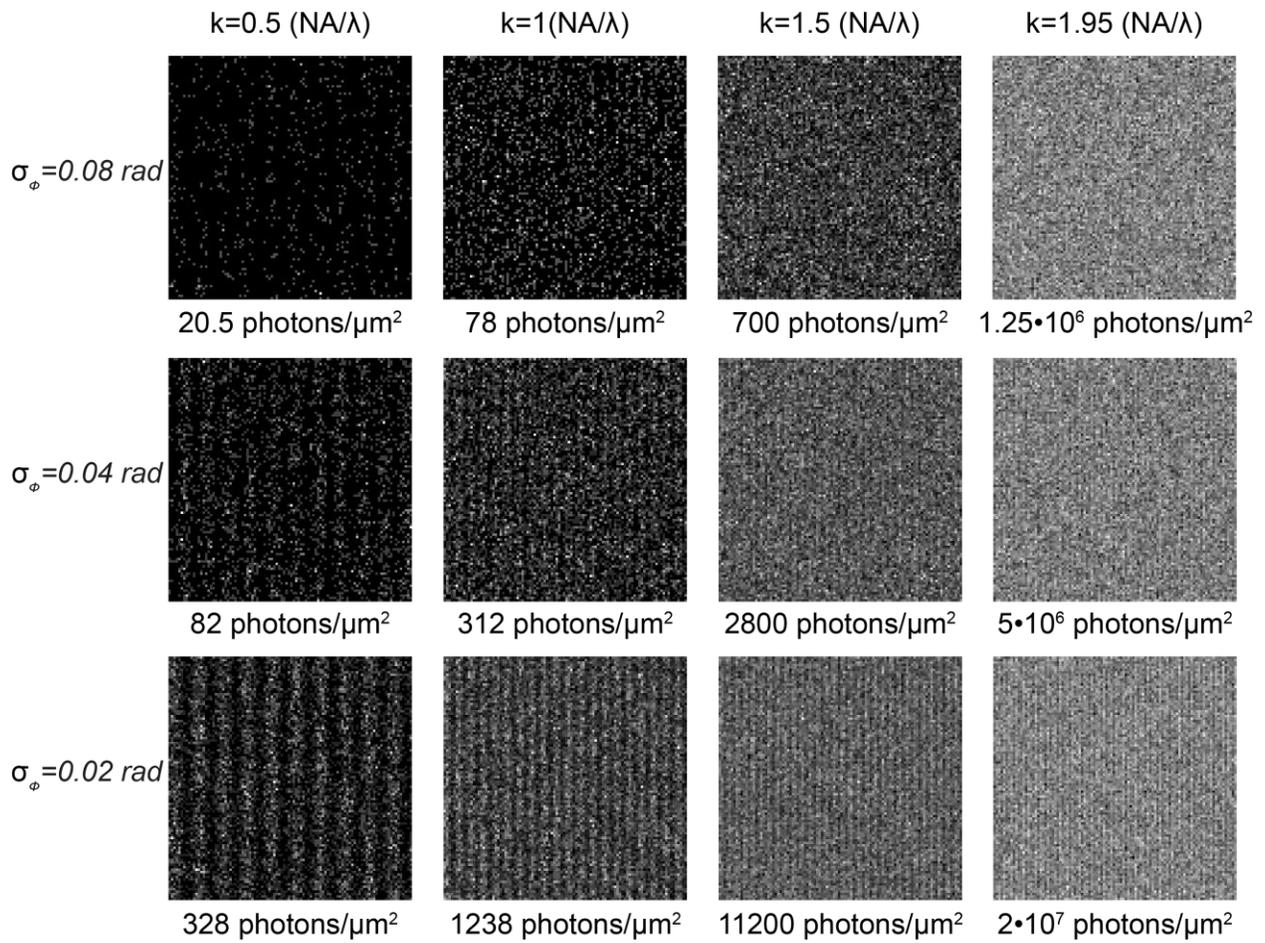

**Fig. S2.** Image visualization in wide-field microscopes with wide range of frequencies, photons and $\sigma_\phi$. Images are simulated under the condition: FOV 10 μm×10 μm, NA=1.4, emission wavelength λ=0.7 μm, immersion media refractive index n=1.5, and camera pixel size 0.1 μm.

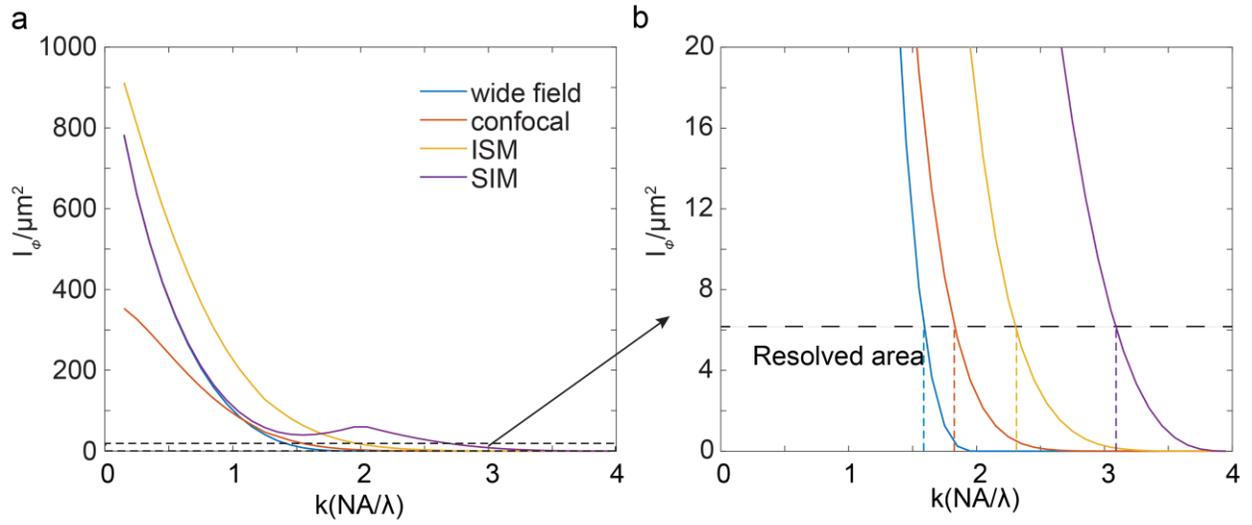

**Fig. S3.** (a) Information $I_\phi/\mu m^2$ calculation in 4 imaging modalities for object of different frequencies. $I_\phi$ is the 3$^{rd}$ row 3$^{rd}$ column term in Fisher information matrix $I(\theta)$ where $\theta = [\alpha, k, \phi]$. (b) Zoom-in plot of (a). Resolved criteria (black dashed line) is defined as $I_\phi/\mu m^2 = 6.25$, which approximately equivalent to $\sigma_\phi = 0.04$ (if consider zero correlation among $[\alpha, k, \phi]$). Color dashed line would be a variant of IbR defined with Fisher information per area. Plots are simulated under the condition: FOV 10 μm×10 μm, NA=1.4, emission wavelength λ=0.7 μm, immersion media refractive index n=1.5, and camera pixel size 0.1 μm. Photon density is 5000 photons/μm$^2$.

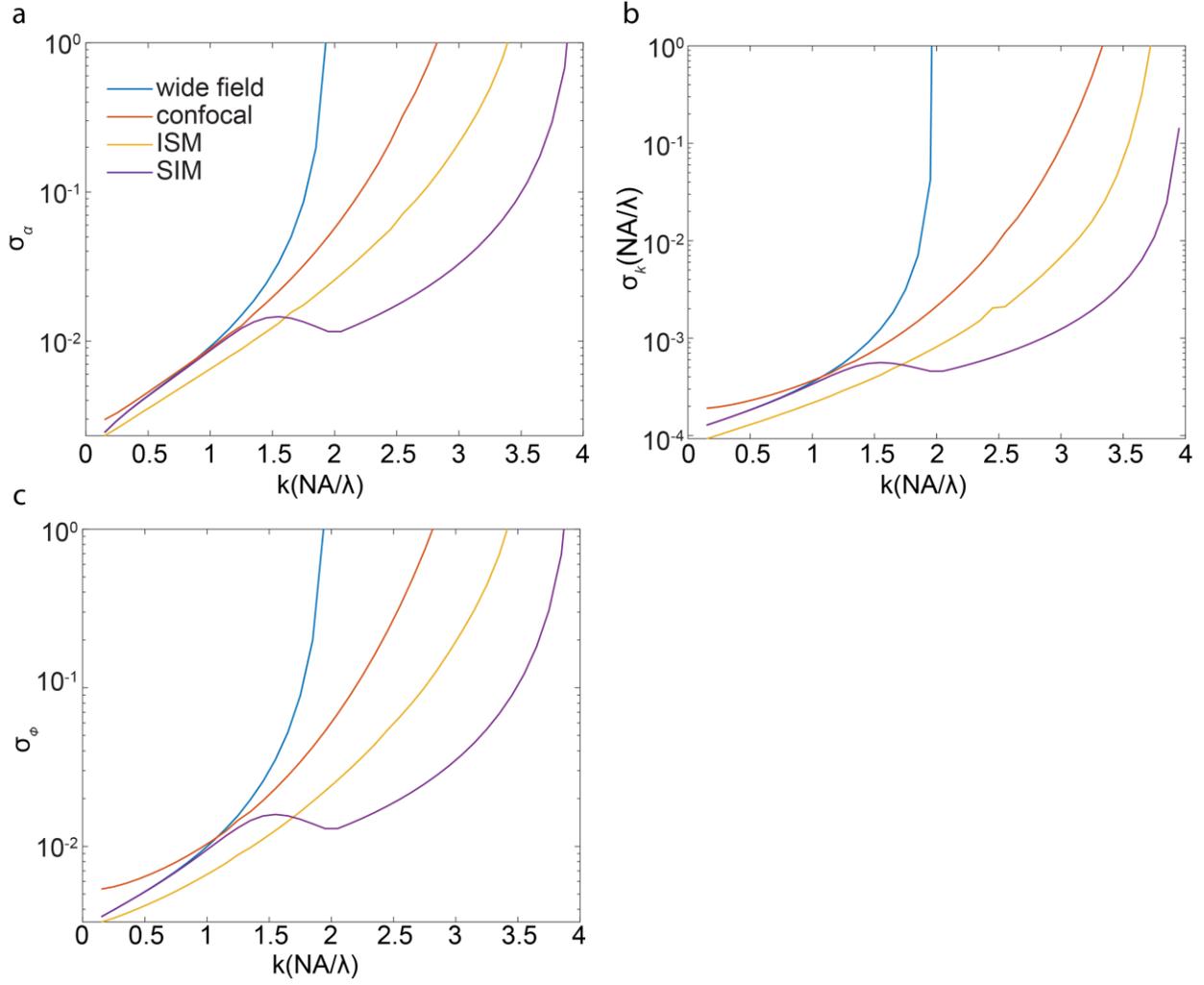

**Fig. S4.** $\sigma$ on all parameters $[\alpha, k, \phi]$ calculated similarly as $\sigma_\phi$. Plots are simulated under the condition: FOV 10 μm×10 μm, NA=1.4, emission wavelength λ=0.7 μm, immersion media refractive index n=1.5, and camera pixel size 0.1 μm. Photon density is 5000 photons/μm$^2$.

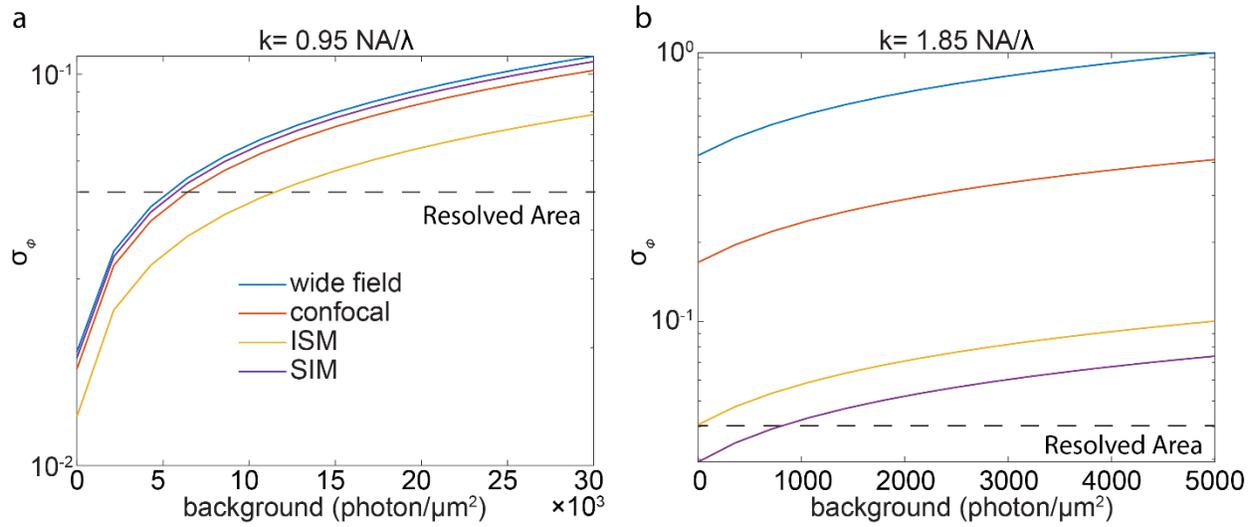

**Fig. S5.** $\sigma_\phi$ with respect to background. Signal photon density is 1000 photons/μm² for objects of different frequencies. Plots are simulated under the condition: FOV 10 μm×10 μm, NA=1.4, emission wavelength λ=0.7 μm, immersion media refractive index n=1.5, and camera pixel size 0.1 μm.

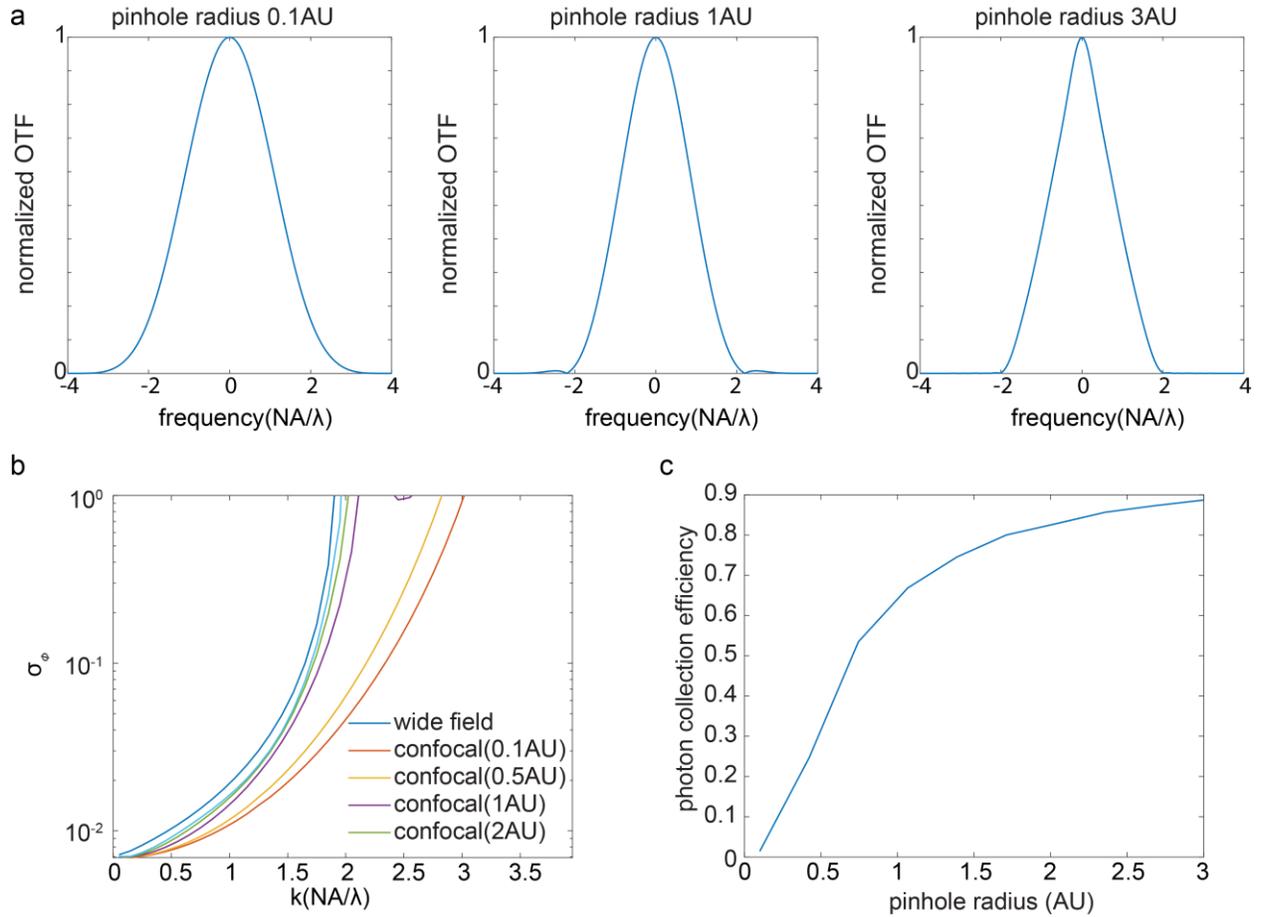

**Fig. S6.** Confocal system's performance, optical transfer function, and photon collection efficiency with respect to its pinhole radius**.** (a) Normalized effective OTF of confocal microscope. (b) $\sigma_\phi$ curve plotted with *<u>fixed collected</u>* photon counts with different pinhole size. Smaller pinhole size provides better resolving power if receiving the same number of photons (c) Photon collection efficiency with respect to pinhole radius. Plots in (b) and (c), are simulated under the condition: FOV 10 μm×10 μm, NA=1.4, emission wavelength λ=0.7 μm, immersion media refractive index n=1.5, and pixel size in wide-field system as well as scanning interval in confocal system 0.1 μm.

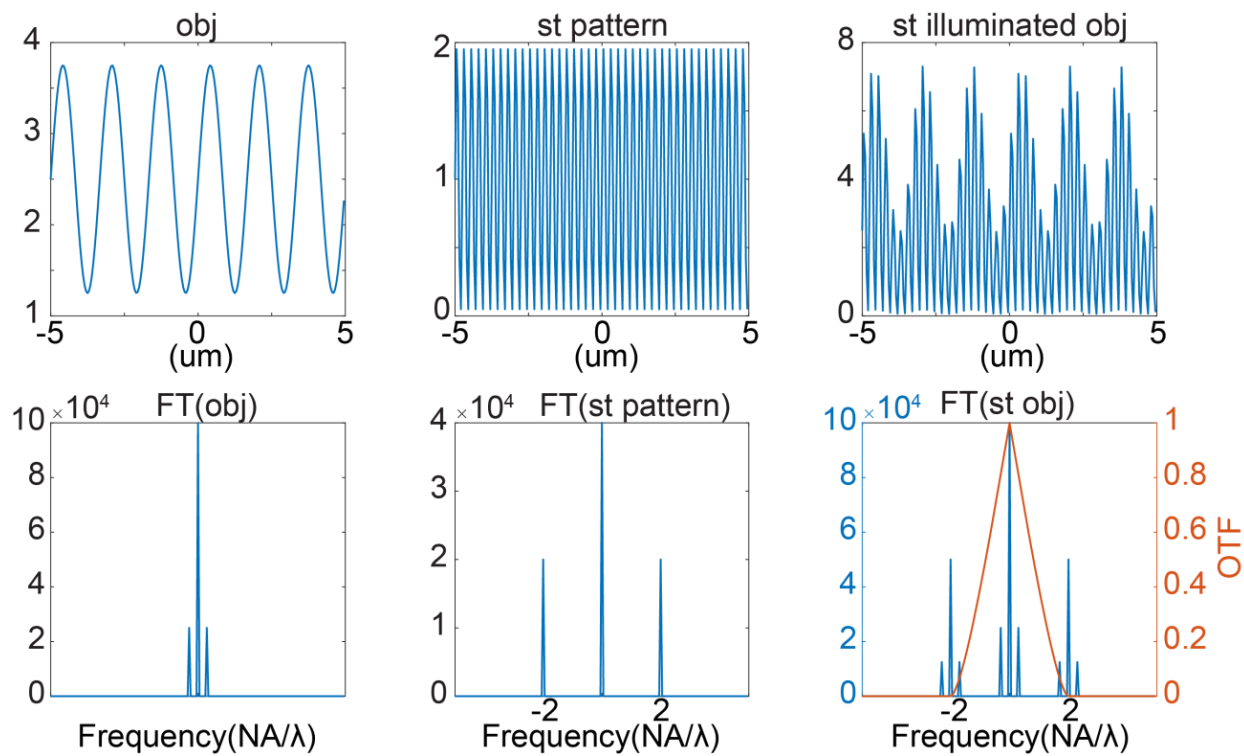

**Fig. S7.** Structured illumination microscope image visualized in 1d when illuminated with lateral sin pattern when object is of low frequency. From left to right, plots are fluorophores distribution, structured illumination pattern intensity distribution and fluorescence distribution of illuminated object. Top plots are in real space while bottom plots are in Fourier space.

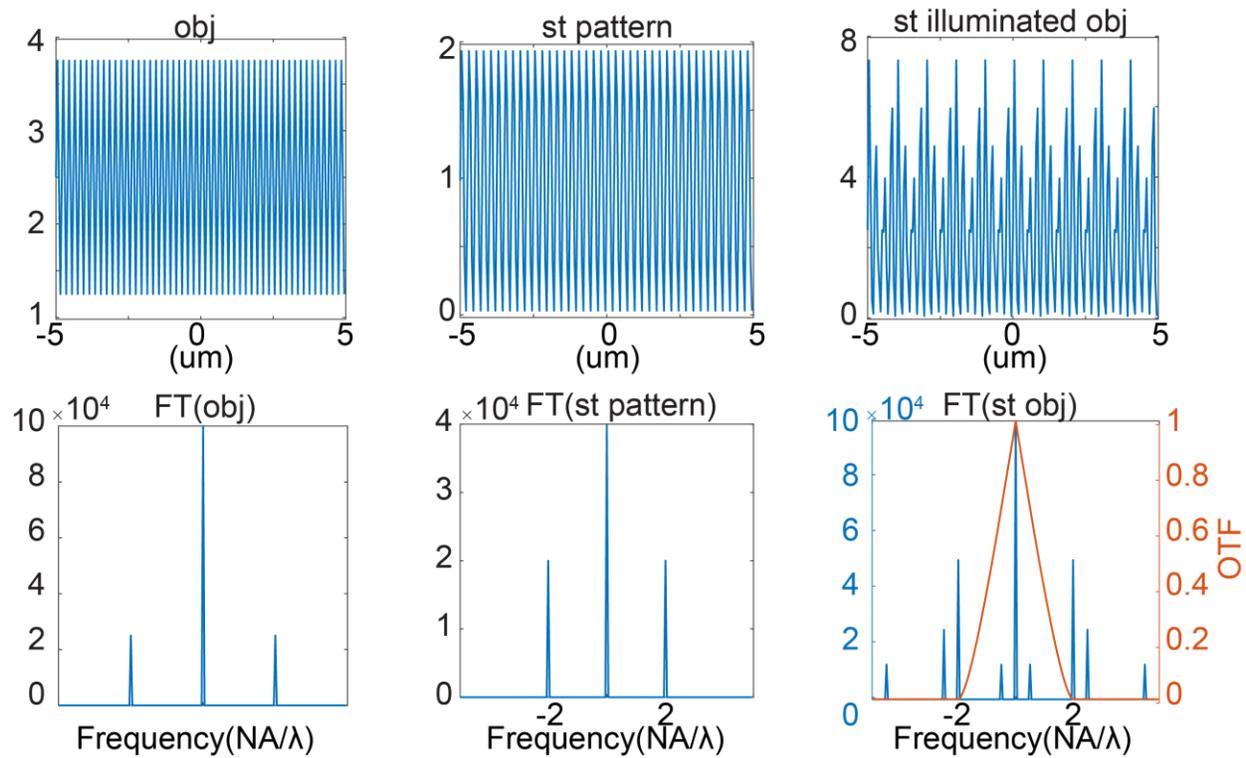

**Fig. S8.** Structured illumination microscope image visualized in 1d when illuminated with lateral sin pattern when object is of high frequency. From left to right, plots are fluorophores distribution, structure illumination pattern intensity distribution and fluorescence distribution of illuminated object. Top plots are in real space while bottom plots are in Fourier space.

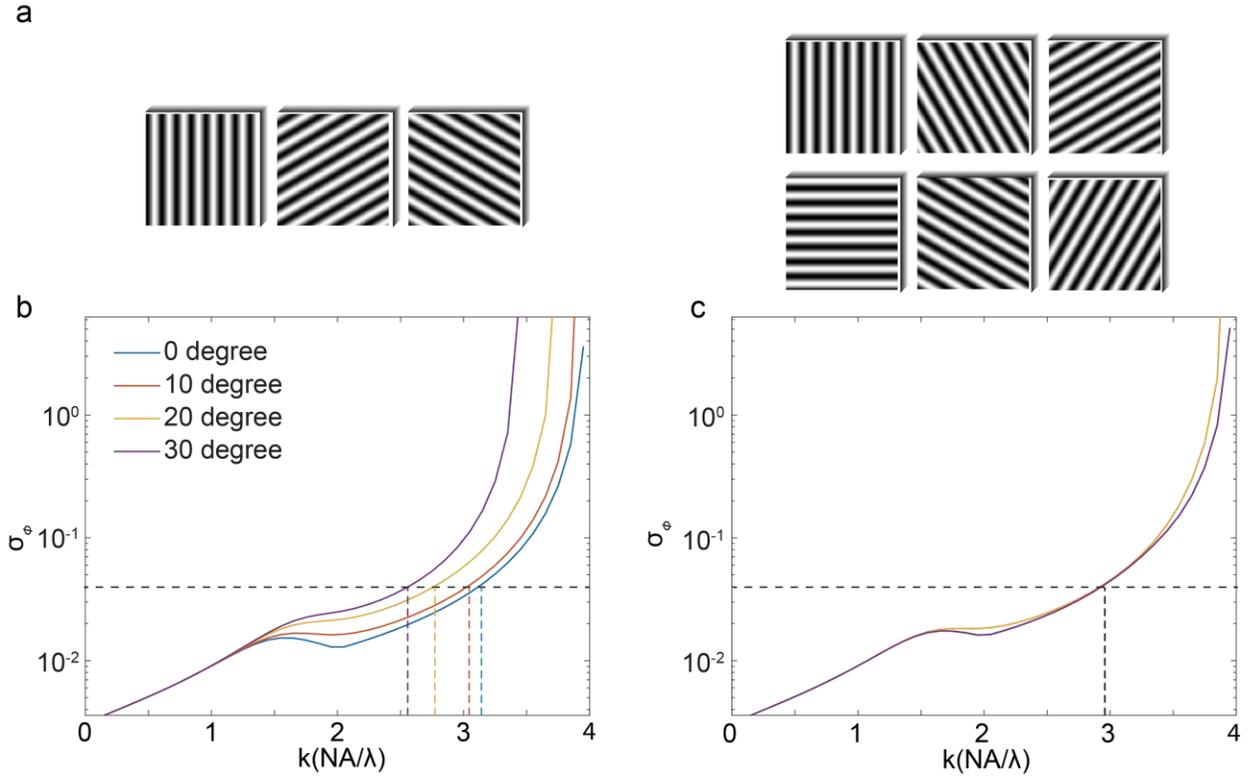

**Fig. S9.** SIM phase estimation precision limit $\sigma_\phi$ with respect to its initial orientation of the structured illumination patterns set. (a) Structured illumination patterns visualization. Each icon (structured illumination pattern of one orientation) containing patterns of 3 phases. (b) $\sigma_\phi$ of SIM with structured illumination patterns set contains 3 orientations. Four initial orientations of illumination patterns set are plotted (angle indicates the degree of a clockwise rotation). (c) $\sigma_\phi$ of with structured illumination patterns set contains 6 orientations. Plots in (b) and (c) are simulated under condition: FOV 10 μm×10 μm, NA=1.4, emission wavelength λ=0.7 μm, immersion media refractive index n=1.5, and camera pixel size 0.1 μm. Photon density is 5000 photons/μm$^2$.

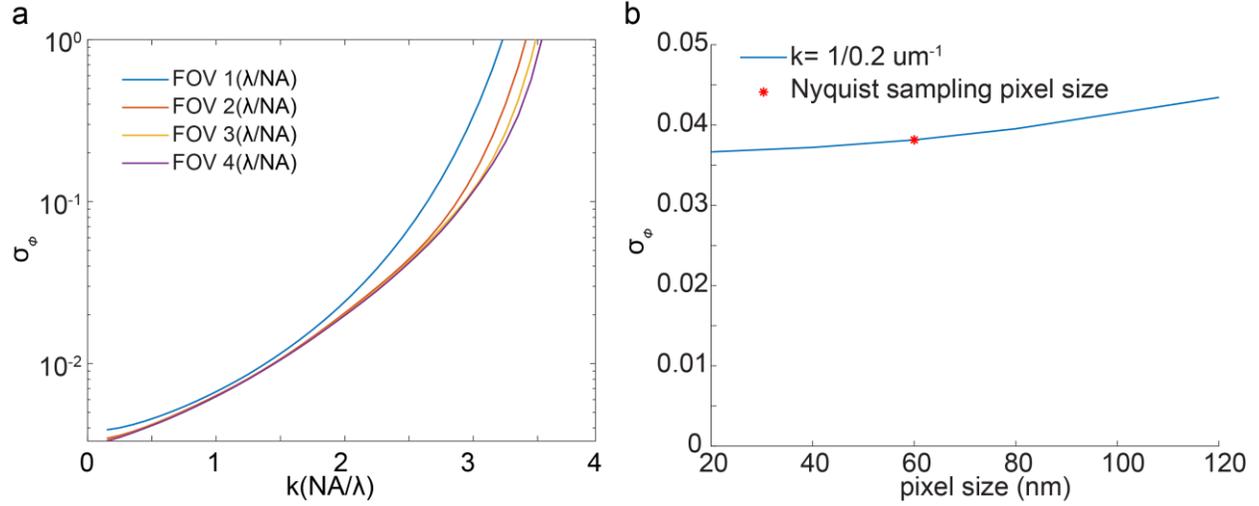

**Fig. S10.** ISM resolving power with respect to FOV of scanning positions and corresponding camera pixel size. (a) $\sigma_\phi$ in ISM with respect to different ISM's field of view size (size of the camera for each scanning position). Camera's FOV greater than $\frac{2\lambda}{NA}$ has little influence among each other. (b) $\sigma_\phi$ in ISM with respect to camera pixel size. FOV is $\frac{4\lambda}{NA}$. The Nyquist sampling pixel size for ISM's highest frequency is $\frac{1}{2*(k_{obj}+k_{illumination})}$. We found the pixel size of camera in ISM have little influence on ISM's resolving power. Plots are simulated under condition: FOV 10 μm×10 μm, NA=1.4, emission wavelength λ=0.7 μm, immersion media refractive index n=1.5. Photon density is 5000 photons/μm².